\documentstyle[aps]{revtex}
\begin{document}
\date{May 8, 2001}
\title{\Large{\bf Instanton method for the electron propagator}}
\author{\large Michael R. Geller}
\address{Department of Physics and Astronomy, University of Georgia, Athens, 
Georgia 30602-2451, USA}
\maketitle

\large

\begin{abstract}\large{
A nonperturbative theory of the electron propagator is developed and used to 
calculate the one-particle Green's function and tunneling density-of-states in
strongly correlated electron systems. The method, which is based on a 
Hubbard-Stratonovich decoupling of the electron-electron interaction combined 
with a cumulant expansion of the resulting noninteracting propagator, provides
a possible generalization of the instanton technique to the calculation of the 
electron propagator in a many-body system. Application to the one-dimensional
electron gas with short-range interaction is discussed.}
\end{abstract}


\section{introduction}

In a large class of one-dimensional systems, electron-electron interaction 
leads to the breakdown of Landau's Fermi liquid theory and to the formation of
a highly correlated Tomonaga-Luttinger liquid phase. In 1990, Wen \cite{Wen 
proposal 1990 PRB} proposed that an edge state in the fractional quantum Hall 
effect (FQHE) regime of the two-dimensional electron gas is a chiral 
Luttinger liquid (CLL), a chiral counterpart to this non-Fermi-liquid state. 
Theoretical work by Wen\cite{Wen reviews} and by Kane and Fisher \cite{Kane 
and Fisher review} showed that the transport and spectral properties of edge 
states in the FQHE regime should be strikingly different than edge states in 
the integral regime, and several of these properties have been observed 
experimentally\cite{cll experiments}.

Recently, however, an important experiment by Grayson {\it et al.}
\cite{Grayson etal} on the tunneling spectra of quantum Hall edges found 
three surprising results: First, power-law tunneling characteristics were 
observed over a range of filling factors $\nu$, irrespective of whether a 
quantized Hall state existed at that filling factor. Second, within a given 
FQHE plateau, the tunneling exponent was found to vary with the filling factor
and was not fixed when the Hall conductance was quantized. Measurements on 
some samples, however, do show a density-of-states (DOS) plateau\cite{Chang 
etal plateau}. And third, the experiment shows that the tunneling DOS 
predicted by the multicomponent CLL theory for the hierarchy states is 
incorrect.

This experiment was motivated by an earlier experiment by Chang {\it et al.} 
\cite {Chang etal 1/2}, who observed CLL-like tunneling characteristics at 
$\nu = 1/2.$ Although some aspects of Ref.~\onlinecite{Grayson etal} are 
beginning to be understood\cite{theoretical work}, especially in connection 
with the third point above, the most fundamental aspects are not.

In this paper I will outline a new theoretical method to calculate the 
one-particle Green's function and tunneling DOS in a strongly correlated 
electron system. The method, discussed in detail in Ref.~\cite{Geller 
instanton}, is based on a Hubbard-Stratonovich decoupling of the 
electron-electron interaction combined with a cumulant expansion of the 
resulting noninteracting propagator. It provides a potential generalization of
the instanton technique\cite{instanton technique} to the calculation of the 
electron propagator in a many-body system \cite{related methods}, and a 
variant of the intuitive but phenomenological ``charge spreading'' picture 
\cite{charge spreading picture} emerges automatically. After describing the 
general formalism I will apply it to spinless fermions with short-range 
interaction. In future work the method will be applied to the sharp and 
smooth edges of FQHE systems.

\section{outline of the method}

I consider a $D$-dimensional interacting electron system, possibly in an 
external magnetic field. For simplicity I will suppress spin indices and
assume the system to be translationally invariant with density $n_0$. The 
grand-canonical Hamiltonian is $ H = H_0 + V$, where
\begin{equation}
V \equiv {\textstyle{1 \over 2}} \int d^{\scriptscriptstyle D}r \ 
d^{\scriptscriptstyle D} r' \, \delta n({\bf r}) \, U({\bf r}-{\bf r}') \, 
\delta n({\bf r}'), \ \ \ \ \
\delta n({\bf r}) \equiv \psi^\dagger({\bf r}) \psi({\bf r}) - n_0.
\label{V}
\end{equation}
$H_0$ is the noninteracting Hamiltonian. We want to calculate the Euclidean 
propagator
\begin{equation}
G({\bf r}_{\rm f},{\bf r}_{\rm i} ,\tau_0) \equiv - \big\langle T \psi_{\! 
\scriptscriptstyle H}({\bf r}_{\rm f} , \tau_0) {\bar \psi}_{\! 
\scriptscriptstyle H}({\bf r}_{\rm i} ,0) \big\rangle_{\! \scriptscriptstyle 
H},
\label{G definition}
\end{equation}
which can be written (in the interaction representation with respect to $H_0$)
as
\begin{equation}
G({\bf r}_{\rm f} , {\bf r}_{\rm i},\tau_0) = - {\big\langle T 
\psi({\bf r}_{\rm f}  ,\tau_0) {\bar \psi} ({\bf r}_{\rm i},0) e^{- 
\int_0^\beta d \tau \, V(\tau)} \big\rangle_0 \over \big\langle T 
e^{- \int_0^\beta d \tau \, V(\tau)} \big\rangle_0 } . 
\end{equation}
A Hubbard-Stratonovich transformation then leads to
\begin{equation}
G({\bf r}_{\rm f}, {\bf r}_{\rm i},\tau_0) = {\cal N} \, {\int D\phi \ 
e^{-{1 \over 2} \int \phi U^{-1} \phi} \ g({\bf r}_{\rm f}, {\bf r}_{\rm i},
\tau_0 | \phi) \over \int D\phi \ e^{-{1 \over 2} \int \phi U^{-1} \phi}} ,   
\label{exact form}
\end{equation}
where
\begin{equation}
g({\bf r}_{\rm f}, {\bf r}_{\rm i}, \tau_0
| \phi) \equiv - \big\langle T \psi({\bf r}_{\rm f} ,\tau_0) 
{\bar \psi}({\bf r}_{\rm i} ,0) \, 
e^{i \int_0^\beta d \tau \int d^{\scriptscriptstyle D}r \, \phi({\bf r},\tau)
\, \delta n({\bf r},\tau) } 
\big\rangle_0 
\label{correlation function definition}
\end{equation}
is a noninteracting correlation function, and ${\cal N} \equiv \langle T 
\exp(-\int_0^\beta d \tau \, V) \rangle_0^{-1}$ is a constant. So far
everything is exact. What remains is to find an appropriate approximation
for $g({\bf r}, {\bf r}', \tau | \phi)$ and to do the resulting functional 
integral.

I evaluate (\ref{correlation function definition}) with a second-order
cumulant expansion,
\begin{equation}
g({\bf r}_{\rm f}, {\bf r}_{\rm i}, \tau_0 | \phi) = G_0({\bf r}_{\rm f}, 
{\bf r}_{\rm i}, \tau_0) \, e^{ \int C_1({\bf r}, \tau) \, \phi({\bf r},\tau) +
\int  C_2({\bf r} \tau, {\bf r}' \tau' ) \, \phi({\bf r},\tau) \, 
\phi({\bf r}', \tau') }.
\end{equation}
The $C_n$ are known in terms of noninteracting Green's functions. For example,
\begin{equation}
C_1({\bf r}, \tau) = -i \, { G_0({\bf r}_{\rm f} ,{\bf r} ,\tau_0 - \tau) \,
G_0({\bf r}, {\bf r}_{\rm i} ,\tau) \over 
G_0({\bf r}_{\rm f} ,{\bf r}_{\rm i} ,\tau_0)}.
\end{equation}
The functional integral in (\ref{exact form}) can be done exactly, leading to
\cite{Geller instanton}
\begin{equation}
G({\bf r}_{\rm f}, {\bf r}_{\rm i},\tau_0) = {\cal A}(\tau_0) \, 
G_0({\bf r}_{\rm f} , {\bf r}_{\rm i},\tau_0) \, e^{-S(\tau_0)}, 
\label{general result}
\end{equation}
where ${\cal A} \equiv {\cal N} \, [ {\rm Det} \, (1-2 C_2 U)]^{-{1 
\over 2}}$ is a fluctuation determinant and
\begin{equation}
S \equiv {\textstyle{1 \over 2}} \int_0^\beta \! d\tau \, d\tau' \! \int \! 
d^{\scriptscriptstyle D}r \, d^{\scriptscriptstyle D}r' \, \rho({\bf r},\tau) 
\, U_{\rm eff}({\bf r} \tau, {\bf r}' \tau') \, \rho({\bf r}',\tau'). 
\label{action}
\end{equation}
Here $\rho({\bf r},\tau) \equiv -i \, C_1({\bf r} \tau)$ and $U_{\rm eff}(
{\bf r} \tau, {\bf r}' \tau') \equiv ( U^{-1} - 2 C_2 )^{-1}_{{\bf r}
\tau, {\bf r}' \tau'}.$

I interpret (\ref{general result}) as follows: $S$ is the Euclidean action for
a time-dependent charge distribution $\rho({\bf r},\tau)$ whose dynamics, 
governed by $H_0$, describes the charge density associated with an electron 
inserted into position ${\bf r} = {\bf r}_{\rm i}$ at time $\tau \! = \! 0$ 
and removed from ${\bf r}_{\rm f}$ at $\tau_0$. The charge interacts via an 
effective interaction $U_{\rm eff}$ that accounts for the modification of the 
electron-electron interaction by dynamic screening. My interpretation of $ -i
\, C_1({\bf r} \tau)$ as the charge density associated with a tunneling 
electron follows from extensive numerical studies and from the exact (at 
$T=0$) identity $\int \! d^{\scriptscriptstyle D}r \, \rho({\bf r},\tau)
= \Theta(\tau) \, \Theta(\tau_0 - \tau).$

\section{application to one-dimensional spinless fermions}

Here I assume a short-range interaction of the form $U(x-x') = U_0 \lambda \,
\Delta(x-x')$, where $\Delta(x)$ is a broadened delta function with range
$\lambda$, and show that (\ref{general result}) correctly predicts a 
power-law DOS. First consider the ``tree-level'' instanton approximation, 
obtained by keeping the first cumulant $C_1$ only. In this case the action is
\begin{equation}
S= {U_0 \lambda \over 2} \int_0^\beta \! d\tau \int_{-\infty}^\infty \! dx \
\big[ \rho(x,\tau) \big]^2,
\label{1D action}
\end{equation}
and we can choose $x_{\rm i} = x_{\rm f} = 0.$ In polar coordinates $x=R \cos
\theta$ and $v_{\rm F} \tau = R \sin \theta$, the low-energy noninteracting 
propagator can be written as
\begin{equation}
G_0(R,\theta) = {\sin(k_{\rm F} R \cos \theta - \theta) \over \pi R},
\label{polar 1D propagator}
\end{equation}
which shows that $G_0$ falls off as $1/R$ in the Euclidean plane ${\vec R} = 
(x, v_{\rm F} \tau)$. Because $\rho(x,\tau)$ has $1/R$ ``singularities'' 
\cite{low energy footnote} at ${\vec R} = (0,0)$ and $(0,v_{\rm F} \tau_0)$, 
the action diverges logarithmically, the infrared divergence from the tail of 
one singularity being cut off by the position of the other, a distance 
$v_{\rm F} \tau_0$ away. $S$ therefore diverges logarithmically in $\tau_0$, 
leading to the well-known power-law DOS in one dimension. In the limit 
$\lambda \ll k_{\rm F}^{-1}$ it can be shown that 
\begin{equation}
S = {3 \over 8 \pi} {U_0 \lambda \over  v_{\rm F}} \ln \bigg({\tau_0 \over 
a}\bigg)  ,
\end{equation}
where $a$ is a microscopic cutoff length, leading to a DOS 
\begin{equation}
N(\epsilon) = {\rm const} \times \epsilon^\delta \ \ \ \ \ {\rm with}
\ \ \ \ \ \delta =  {3 \over 8 \pi} {U_0 \lambda \over  v_{\rm F}}.  
\end{equation}

Including the second cumulant $C_2$ in does not qualitatively change this 
picture. However, it affects the value of the DOS exponent $\delta$. Although 
the ultimate limitations of this method are not understood at present, 
application of the second-cumulant (or ``one-loop'') analysis to the spinless 
Tomonaga-Luttinger model leads to an exponent $\delta$ in {\it exact} 
agreement with the bosonization result
\cite{Geller instanton}.

\section{acknowledgements}

This work was supported by the National Science Foundation under Grant 
No.~DMR-0093217, and by a Cottrell Scholars Award from the Research 
Corporation. 
It is a pleasure to thank
Claudio Chamon, 
Yong Baek Kim,
Allan MacDonald,
Gerry Mahan, 
David Thouless,
and Giovanni Vignale 
for useful discussions.


\begin{references}
\bibitem{low energy footnote} These singularities arise from using the 
low-energy approximation (\ref{polar 1D propagator}) to the noninteracting 
propagator; they are not present in the exact expression. However, the 
$\tau_0$-dependent part of $S$ comes from the $1/R$ tail of $G_0$, which is 
correctly described by the low-energy form.
\end{references}
\end{document}